\documentclass[%
    twocolumn,
 superscriptaddress,
 amsmath,amssymb,
 aps, showpacs, 
 prper,
floatfix,
dvipsnames,
]{revtex4-2}

\usepackage{dcolumn}
\usepackage{tikz}
\usepackage{scalefnt}
\usepackage{braket}
\usepackage{derivative}
\usepackage{hyperref}
\usepackage{optidef}
\usepackage{graphicx}
\usepackage{dcolumn}
\usepackage{bm}
\usepackage{dsfont}
\usepackage{mathrsfs}
\usepackage{mathtools}
\usepackage{multirow}
\usepackage[caption=false]{subfig}
\usepackage{algpseudocode}
\usepackage{algorithm}
\usepackage{orcidlink}

\DeclareMathOperator{\im}{\mathring \imath}
\usepackage{physics}

\begin{document}

\newcommand{\ms}[1]{\mbox{\scriptsize #1}}
\newcommand{\msb}[1]{\mbox{\scriptsize $\mathbf{#1}$}}
\newcommand{\msi}[1]{\mbox{\scriptsize\textit{#1}}}
\newcommand{\nn}{\nonumber} 
\newcommand{\dg}{^\dagger}
\newcommand{\smallfrac}[2]{\mbox{$\frac{#1}{#2}$}}
\newcommand{\pfpx}[2]{\frac{\partial #1}{\partial #2}}
\newcommand{\dfdx}[2]{\frac{d #1}{d #2}}
\newcommand{\half}{\smallfrac{1}{2}}
\newcommand{\s}{{\mathcal S}}
\newcommand{\rev}{\color{RedOrange}}
\newtheorem{theo}{Theorem} \newtheorem{lemma}{Lemma}

\title{Globally optimal control of quantum dynamics}

\author{Denys I. Bondar\orcidlink{0000-0002-3626-4804}}
 \email{dbondar@tulane.edu}
\affiliation{%
 Department of Physics and Engineering Physics, Tulane University,  New Orleans, LA 70118, USA
}%

\author{Llorenç Balada Gaggioli\orcidlink{0009-0006-3930-3457}}
\email{llorenc.balada.gaggioli@fel.cvut.cz}
\affiliation{%
Czech Technical University in Prague, Prague 121 35, Czech Republic}
\affiliation{%
LAAS-CNRS, Université de Toulouse, France
}%

\author{Georgios Korpas\orcidlink{0000-0003-3850-4979}}
\email{georgios.korpas@hsbc.com} 
\affiliation{%
Quantum Technologies Group, HSBC Lab, 20 Pasir Panjang Road, 117439, Singapore}
\affiliation{%
Czech Technical University in Prague, Prague 121 35, Czech Republic}
\affiliation{%
Archimedes Research Unit on AI, Data Science and Algorithms, Athena Research Center, 1 Artemidos Street
15125, Marousi, Greece }

\author{Jakub Marecek\orcidlink{0000-0003-0839-0691}}
\email{jakub.marecek@fel.cvut.cz}
\affiliation{%
Czech Technical University in Prague, Prague 121 35, Czech Republic}

\author{Jiri Vala\orcidlink{0000-0001-7795-8602}}
\email{jiri.vala@mu.ie}
\affiliation{Department of Physics, Maynooth University, Ireland}
\affiliation{School of Theoretical Physics, Dublin Institute for Advanced Studies, Dublin, Ireland}
\affiliation{Tyndall National Institute, Cork, Ireland}

\author{Kurt Jacobs\orcidlink{0000-0003-0828-6421}}
\email{dr.kurt.jacobs@gmail.com}
\affiliation{US DEVCOM Army Research Laboratory, Adelphi, Maryland 20783, USA}
\affiliation{Department of Physics, University of Massachusetts at Boston, Boston, Massachusetts 02125, USA}

\date{\today}

\begin{abstract}
Optimization of constrained quantum control problems powers quantum technologies. This task becomes very difficult when these control problems are nonconvex and plagued with dense local extrema. For such problems current optimization methods must be repeated many times to find good solutions, each time requiring many simulations of the system. Here, we present \emph{quantum control via polynomial optimization} (QCPOP), a method that eliminates this problem by directly finding globally optimal solutions. The resulting increase in speed, which can be a thousandfold or more, makes it possible to solve problems that were previously intractable. This remarkable advance is due to global optimization methods recently developed for polynomial functions. We demonstrate the power of this method by showing that it obtains an optimal solution in a single run for a problem in which local extrema are so dense that gradient methods require thousands of runs to reach a similar fidelity. 
Since QCPOP is able to find the global optimum for quantum control, we expect that it will not only enhance the utility of quantum control by making it much easier to find the necessary protocols, but also provide a key tool for understanding the precise limits of quantum technologies. Finally, we note that the ability to cast quantum control as polynomial optimization resolves an open question regarding the computability of exact solutions to quantum control problems. 
\end{abstract} 

\maketitle 

\section{Introduction}

Controlling the evolution of microscopic quantum systems is essential for the development of future quantum technologies~\cite{Palao_02, Nielsen_06, Roque_21, Motzoi_09,Atalaya2021,berg2022probabilistic}. The primary method of implementing such control  is to change the Hamiltonian of the system with time~\cite{Koch_22, morzhin_krotov_2019, zhang_quantum_2017, glaser2015training, petersen_quantum_2010, brif_control_2010, dalessandro_introduction_2007}. This change can be implemented by changing classical control fields, although the number of these ``control knobs'' is usually very limited. Fortunately, the algebra of linear operators is sufficiently complex that even changing the size of a single term in the Hamiltonian, given a sufficient time, will usually provide universal control of a system. The value(s) of the control field(s) as a function of time are referred to as the \textit{control functions} or simply \textit{controls}. 

It has been observed that in the absence of significant constraints, finding controls to generate a particular final state or a unitary operation with high fidelity can be achieved easily using standard gradient search methods~\cite{rabitz_quantum_2004, brif_control_2010}. This implies that the search ``landscape" in the space of controls is largely free of local extrema. However, constraints are usually an important part of control problems for applications in quantum technology. The most ubiquitous is a constraint on the duration of the control; given ever-present dissipation and decoherence processes, it is usually important to achieve the goal in the shortest possible time. Specific classes of control problems require other kinds of constraints, examples of which are hybrid quantum–classical optimization and variational quantum algorithms~\cite{magann_pulses_2021,ge_optimization_2022}. 

The methods most commonly employed to solve quantum control problems use gradient search methods. The simplest approach is to discretize the control function in some way and use a state-of-the-art gradient search method such as the Broyden-Fletcher-Goldfarb-Shanno algorithm~\cite{Nocedal06}. Alternatively, a number of methods that add some additional sophistication prior to applying a gradient search are commonly used: GRAPE (GRadient Ascent Pulse Engineering)~\cite{Khaneja2005,PhysRevA.84.022305,de2011second,heeres2017implementing}, CRAB (Chopped RAndom-Basis)~\cite{doria_optimal_2011, caneva_chopped_2011, rach_dressing_2015}, and the Krotov method~\cite{Krotov_83, Goerz_2018,  morzhin_krotov_2019}. For control problems with a significant density of local extrema (``traps''), these methods must be run many times, and if the density is too high, they simply fail to find good solutions. Stochastic search algorithms~\cite{judson_teaching_1992} such as simulated annealing, a plethora of reinforcement and machine learning methods~\cite{bukov_reinforcement_2018, palittapongarnpim_learning_2017, niu_universal_2019, Dalgaard_2020}, and a Bayesian approach \cite{mukherjee_bayesian_2020} have all been employed to address this issue. While these machine learning search methods have achieved some significant improvements over gradient methods, fundamentally all of them merely search for local optima, and thus must cover a significant fraction of the search space to raise the likelihood of finding a global optimum. 

In the last few years, tremendous advances have been made in solving optimization problems for polynomials (see Appendix~\ref{sec:PolyOptTheory} for a review). The methods are remarkable in their ability to bypass local optima and locate the global optimum. These methods have two stages. In the first stage, they locate a global optimum, and in particular are able to \textit{guarantee} that the optimum is a global optimum. In the second stage the methods find an argument to the function that achieves this optimum. The second stage is not yet as robust as the first stage, but the implementations we employ have been able to find arguments that realize the global optimum in all the problems we have tried. 

To apply polynomial optimization methods to quantum control we develop a procedure to approximate quantum control problems with polynomial optimization problems. Note that even with such a procedure it is not guaranteed that polynomial optimization will provide a tremendous increase in speed and optimality over traditional search methods: It could be that to achieve a sufficiently good polynomial approximation results in a polynomial optimization problem that is so complex as to require very large numerical resources even for the powerful algorithms that are able to locate the global optima. Here we demonstrate, at least for the common nontrivial examples we consider, that this is not the case; not only does a single run of a polynomial optimization algorithm take a time similar to that of state-of-the-art search methods, but with that single run it finds an optimum solution whereas search methods may require thousands of runs if they are able to find good solutions at all. Not only does this tremendously increase the speed at which quantum control protocols can be obtained, but makes it possible to solve problems that were previously infeasible. As we show, they are also able to find optimal solutions where previous methods could not.

\section{Background: Quantum Control}

Consider a finite-level quantum system described by a Hilbert space $\mathcal{H}$. Both a quantum state $\ket{\psi}$ and the unitary operator, $U(T)$, that evolves the system to time $T$ evolve under Schr\"odinger equation:
\begin{align}\label{Eq:Schrodinger}
    \partial_t \ket{\psi(t)} & = A(t)\ket{\psi(t)} , \;\;\;\ket{\psi(0)}  = \ket{\psi_0} , \\
    \partial_t U(t) & = A(t)U(t), \;\;\; U(0) = \mathds{1} . \label{EqPropagator}
\end{align}
Here $A(t)$ is anti-Hermitian and given by $A(t) = -\im H$, where $H$ is the Hamiltonian of the system. (We have set $\hbar =1$ for convenience.) To control the system, we change one (or more) terms in the Hamiltonian with time. Therefore, we set $H = H_0 + E(t)V$, where $E(t)$ is the control function. The constant part of the Hamiltonian, $H_0$, is called the drift and the time-varying part, $V$ is called the control Hamiltonian. 

The vast majority of control problems for quantum systems are of the following four kinds. The first is called the \emph{fixed-time} or \emph{finite-horizon} control, and involves finding the control $E(t)$ that generates (as near as possible) a ``target" unitary, $U^\star$, at the end of a given evolution time $T$. Specifically,  
\begin{mini}|l|
{E(t)}{ F[U(T),U^\star]}{}{}
\addConstraint{\partial_t U(t) = A(t)U(t), \;\; U(0) = \mathds{1}}{}{}
\addConstraint{A(t) = -\im (H_0 + E(t) V)} . 
\label{eq:QuantumCoherentControl}
\end{mini}
where $F$ is a function whose minimum is at $U = U^\star$. Most forms for $F$ that are typically used are compatible with formulating a polynomial optimization problem. Two common forms are $F = \left\| U(T) - U^\star \right\|^2$ and $F = 1- \left| \Tr[U(T)^{\dagger} U^\star] \right|/N$ \cite{basilewitsch_quantum_2019}. Here $N$ is the dimension of the system and $\left\| M \right\| \equiv \sqrt{\mbox{Tr}[M^\dagger M]}$ is the Hilbert-Schmidt norm of an operator, $M$ (also called the Frobenius or $L_{2,2}$ norm). The second form we will refer to as infidelity, since it is unity minus the fidelity of $U(T)$ with respect to $U^\star$~\cite{NielsenChuang}. 

The second type of control problem is identical to Eq.~(\ref{eq:QuantumCoherentControl}), except that instead of using the evolution operator $U(t)$, the wave function formulation $|\psi(t)\rangle$ can be employed with the specified initial condition $|\psi(0)\rangle = |\psi_0\rangle$. In this case, rather than achieving a specific final state $|\psi^*\rangle$ at final time $T$, we might want to maximize some property of the final state, such as the expectation value of an operator $\hat{O}$. The objective function would then be chosen as $F(\hat{O},|\psi^*\rangle) = -\langle\psi^*|\hat{O}|\psi^*\rangle$.

The third kind of control problem is the minimum-time control. In this case, we still want to achieve a target unitary, but we want to do so in the minimum time. The statement of this problem is 
\begin{mini}|l|
{E(t)}{T}{}{}
\addConstraint{T \geq 0, \;\; F[U(T),U^\star]\leq \varepsilon^2}{}{}
\addConstraint{\partial_t U(t) = A(t)U(t), \;\; U(0) = \mathds{1}}{}{} 
\addConstraint{A(t) = -\im (H_0 + E(t) V)} .
\label{eq:MinTimQuantumControl}
\end{mini}
The fourth and final problem we include here is the same as Eq.~(\ref{eq:MinTimQuantumControl}), but with the unitaries replaced by quantum states. This minimum-time problem for state evolution is the quantum version of the brachistochrone problem~\cite{XWang2015, kuzmak_quantum_2015}

While in the rest of the paper we focus on the single control, for completeness, let us state the generalization of problem~\eqref{eq:QuantumCoherentControl} to the case of multiple controls
\begin{mini}|l|
{\{E_k(t)\}_{k=1}^c}{ F[U(T),U^\star]}{}{}
\addConstraint{\partial_t U(t) = A(t)U(t), \;\; U(0) = \mathds{1}}{}{}
\addConstraint{A(t) = -\im \left(H_0 + \sum_{k=1}^c E_k(t) V_k\right)} . 
\end{mini}

\section{The QCPOP method} 

We obtain a polynomial formulation of our four control problems using three steps. First, we employ the celebrated Magnus expansion to $U(T)$~\cite{Magnus1954}: 
\begin{align}\label{eq:MagnusApprox}
    U(T) = \lim_{n \rightarrow \infty} \exp [ \Lambda_n ], 
    \qquad \Lambda_n = \sum_{k=1}^{n} \Omega_k . 
\end{align}
Using $B(s,t) \equiv [A(s), A(t)]$, $C(s,t,u) \equiv [A(s), B(t,u)]$ the first three terms of the Magnus expansion are 
\begin{align}
    \Omega_1 &= \int_0^T \!\! A(t) \, dt  , \label{eq:Omega1} \;\;\; \Omega_2 = \frac{1}{2} \int_0^T \!\! \int_0^{t} \!\! B(t,s) \, dt  ds , \\
    \Omega_3 &= \frac{1}{3!} \int_0^T \!\! \int_0^{t} \!\! \int_0^{s} \!\!  [C(t,s,u)  -  C(u,t,s)]  \, dt ds du. \label{eq:Omega3}
\end{align}
The Magnus series converges so long as $ 
    \int _{0}^{T}\| A(t) \|\, dt < \pi$. 
An important feature of the Magnus expansion is that  $\exp [ \Lambda_n ]$ is unitary for every $n$. Since the Frobenius norm is unitary invariant, the objective function in Eq.~\eqref{eq:QuantumCoherentControl} can now be approximated as 
\begin{align}\label{Eq:UminusUtarget}
  F[U(T),U^\star] = \left\| U(T) - U^\star \right\|^2 \approx \left\| e^{\Lambda_n/2} -  e^{-\Lambda_n/2} U^\star \right\|^2.
\end{align}
Our second step is to use the Chebyshev polynomial expansion for the matrix exponent~\cite{Tal-Ezer_84}: 
\begin{align}
 e^{\alpha \Lambda_n} & \approx J_0(\alpha) \mathds{1} + 2 \sum_{k=1}^p J_k(\alpha) T_k, 
    \label{Chebapp}
\end{align}
where $J_k(x)$ is the Bessel function and $T_k$ is the matrix-valued Chebyshev polynomial defined via the recurrence relation $T_{k + 1} = 2\Lambda_n T_k + T_{k-1}$, with $T_0 = \mathds{1}$ and $T_1 = \Lambda_n$. Note that the Chebyshev polynomials form an especially efficient basis for the interpolation due to how they distribute approximation errors: Instead of allowing large errors in some regions while achieving perfect accuracy elsewhere, the Chebyshev polynomials spread the error evenly across the entire interpolation interval. This ``equioscillation'' property means the maximum error is minimized, making them nearly optimal for polynomial approximation.

The final step is to use a series representation for the control function, $E(t)$, truncated to a fixed number of terms. Expansion of the control in \emph{any} basis $\{ f_k(t) \}_{k=0}^{m-1}$ will suffice, i.e., 
\begin{align}
    E(t) = \sum_{k=0}^{m-1} x_k f_k(t).    
\end{align}
In our examples below, we choose a power series so that 
\begin{align}\label{eq:E2x}
    E(t) = \sum_{k=0}^{m-1} x_k t^{k} . 
\end{align}
Finding the optimal control function, therefore, involves finding the optimal values of the coefficients $x_k$. All four of our control problems will be polynomial optimization problems so long as $F[U(T),U^\star]$ is a polynomial in $x_k$. Substituting the series representation for $E(t)$ into $A(t)$ and the latter into the integrals for the terms $\Omega_n$ in the Magnus expansion, we see that these are polynomials in $x_k$ and so are $\Lambda_n$. Using the Chebyshev expansion for the Magnus approximation to $U(T)$, Eq.~\eqref{Chebapp}, then gives us $U(T)$ as a polynomial in $x_k$. Finally, since $\left\| U(T) - U^\star \right\|^2$ is a polynomial of the elements of $U(T)$, our four control problems are now polynomial optimization problems. 

The chain of approximations (expansions) we have used to write our control problems as polynomial optimizations is characterized by the order at which we truncate each of the three expansions. Thus, an implementation of QCPOP is characterized by the tuple $(n,p,m)$, where $n$ is the number of terms used in the Magnus expansion, $p$ the number of terms in the Chebyshev expansion, and $m$ the number of terms in the control function. Using problems from Secs.~\ref{SecEvaluatingQCPOP} and~\ref{SecHI} as benchmarks, we have tested the performance of QCPOP for different values of $n, p$. We have empirically found that $n=3$ and $p=5$ (see Appendices~\ref{sec:IllustrationCoherentContr} and \ref{sec:HamiltonianIdentification} for details) are the optimal combinations, meaning that using larger values does not yield any improvement.

\subsection{Modification for piecewise-constant control}

If we want a control function that is piecewise constant, meaning that it consists of contiguous segments with a (possibly different) constant value on each, then the polynomial approximation simplifies. In this case, we do not need the Magnus expansion. If we have $m$ segments in which the $j^{\rm{th}}$ segment has duration $t_j$, then the unitary evolution is given by $U(T) = \prod_{j=1}^m U_j$, where 
\begin{align}
    U_j = e^{-\im (H_0 + E_j V) t_j} .  
\end{align}
Here, the $m$ values $E_j$ are the parameters to be optimized. This time we use the Chebyshev expansion, Eq.~(\ref{Chebapp}), for each of the unitaries $U_j$. Substituting $U(T) = \prod_{j=1}^m U_j$ into the objective function $F[U(T),U^\star]$ results in a polynomial optimization problem. 
 
\subsection{Polynomial optimization} 

As mentioned above, the advantage of formulating quantum control as polynomial optimization is the remarkably powerful global optimization methods for this task that have been developed in the last several years. Here, we use two of these methods, which are already available as libraries in Julia. The first is called the ``moment sum-of-squares hierarchy method'' (for which the Julia library is called TSSOS~\cite{TSSOS,ChordalTSSOS}) and the second is the ``homotopy continuation method'' (for which the Julia library is HomotopyContinuation~\cite{HomotopyContinuation.jl}). In Appendix~\ref{sec:PolyOptTheory}, we give a pedagogical introduction to both the moment sum-of-squares hierarchy method and the homotopy continuation method. Empirically, we have found that homotopy continuation is better when minimizing infidelity ($F = 1- \left| \Tr[U(T)^{\dagger} U^\star] \right|/N$) and the sum-of-squares method is better when minimizing the Hilbert-Schmidt distance ($F = \left\| U(T) - U^\star \right\|^2$).  

\begin{figure}
    \leavevmode\includegraphics[width = 1\columnwidth]{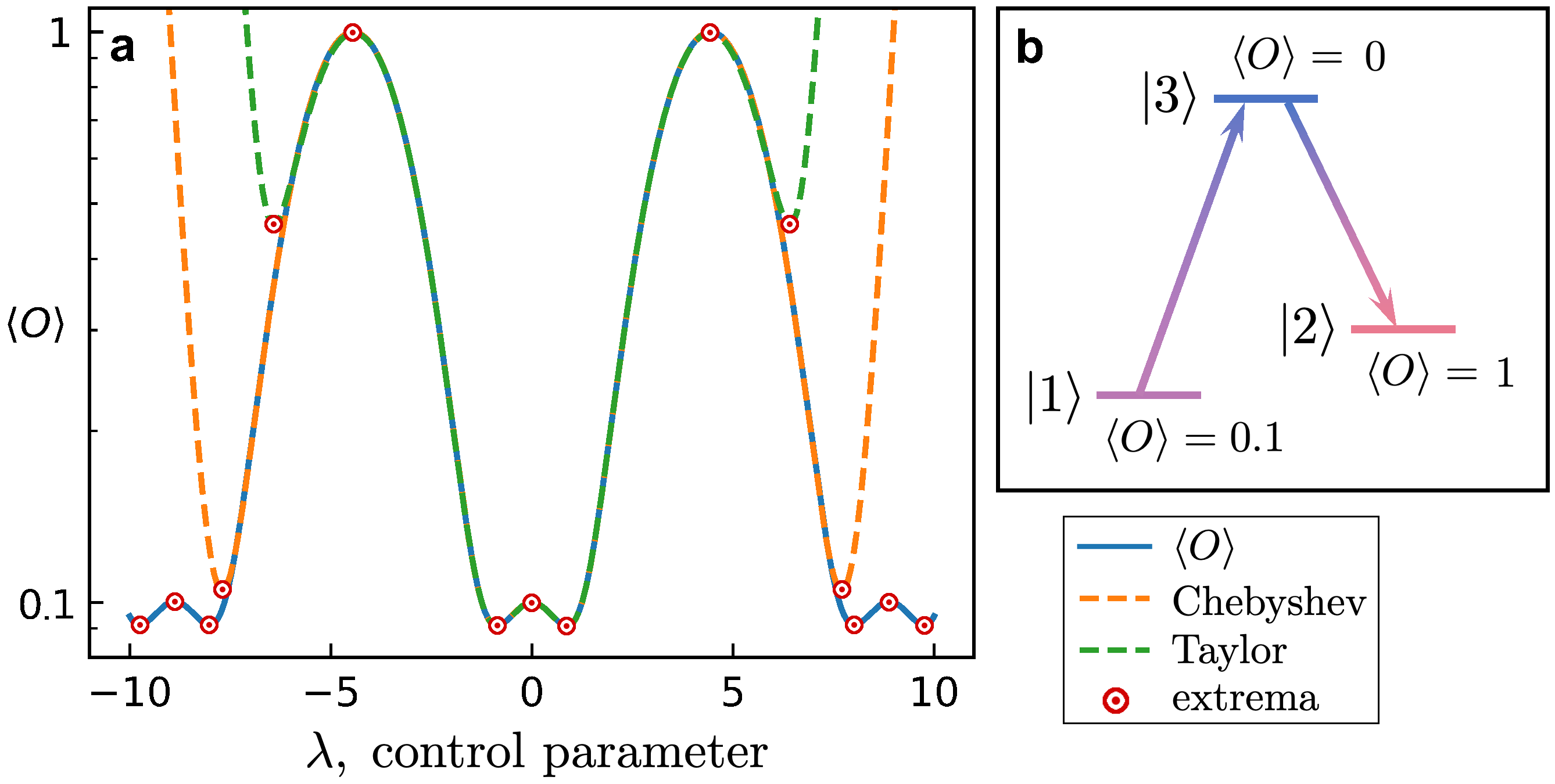}
    \caption{The control problem of Pechen and Tannor~\cite{pechen_are_2011} which has a trap at the point where the control variable, $\lambda$, vanishes. Using QCPOP the trap poses no problem for the homotopy continuation method which returns all the extrema (shown by the red crosses). a) The solid blue line is the value of the objective function (the expectation value of $O$), being a symmetric function of the control variable, $\lambda$. The green (orange) dashed line is the objective function resulting from using the Taylor (Chebyshev) approximation in QCPOP (to tenth order). b) A diagram of the control problem. The system has three levels, each with a different value of the observable, $O$. The control Hamiltonian, $V$, couples $\ket{1}$ to $\ket{2}$ and $\ket{2}$ to $\ket{3}$. }
    \label{fig:OwithTrap}
\end{figure}

\section{Evaluating the Performance of QCPOP}\label{SecEvaluatingQCPOP} 

We first apply QCPOP to a special and simple example in which the global maxima and local traps of the objective function are known. We do this to demonstrate without any ambiguity that the optimization algorithm locates global maxima regardless of the presence of traps. The example we use was originally introduced by Pechen and Tannor in~\cite{pechen_are_2011}, and involves a three-level system that is depicted in Fig.\ref{fig:OwithTrap}b. The drift Hamiltonian, $H_0$, has eigenstates $\ket{1}$, $\ket{2}$, and $\ket{3}$, and the system starts in state $\ket{1}$. The goal is to maximize the expectation value of an observable, $O$, after an evolution time, $T=1$, where the values of $O$ for each of the states $\ket{1}$, $\ket{2}$, and $\ket{3}$ are, respectively, $0.1$, $0$, and $1$. Since the system starts in state $\ket{1}$, achieving the goal requires turning on the control Hamiltonian to transfer the system to state $\ket{3}$. The control Hamiltonian, $V$, does not couple $\ket{1}$ directly to $\ket{2}$, but couples $\ket{1}$ to $\ket{3}$ and $\ket{3}$ to $\ket{2}$ as shown in Fig.\ref{fig:OwithTrap}b. (In particular, all elements of $V$ are zero except $v_{13}=v_{23}=v_{31}=v_{32}=1$.) The Hamiltonian is $H = H_0 + \lambda V$, and we fix $\lambda$ to be constant over the full evolution from $0$ to $1$. We therefore have just a single control variable (the value of $\lambda$). 

The reason why the above control problem has a local maximum (trap) at $\lambda=0$ is quite simple. Since the control must take the system from $\ket{1}$ to $\ket{3}$ via $\ket{2}$, for small values of $\lambda$ the system has a higher probability of being  in $\ket{2}$ than in $\ket{3}$, thus \textit{reducing} the expectation value of $O$ below that of $\ket{1}$. The objective function thus initially decreases as $|\lambda|$ moves away from zero before it increases to the global optimum at $\ket{3}$, hence producing the local maxima at $\lambda = 0$. In Fig.~\ref{fig:OwithTrap} we plot the objective as a function of $|\lambda|$ as well as the approximation to the objective produced by QCPOP with the Chebyshev polynomial truncated at $p=10$. The homotopy continuation optimization method applied to QCPOP finds \textit{all} the extrema and is therefore unaffected by the presence of local maxima~\footnote{Details are given in the Jupyter notebook \url{https://github.com/dibondar/QControlPolyOpt/blob/master/Three_level_system_with_trap--piecewise_constant.ipynb}}. Gradient search methods, if started at or near zero, find only the local maximum. 

We now test QCPOP on a set of nontrivial problems of the kind that quantum control is often employed to solve, and compare it directly to two state-of-the-art methods, GRAPE~\cite{Khaneja2005,PhysRevA.84.022305,de2011second,heeres2017implementing} and CRAB~\cite{doria_optimal_2011, caneva_chopped_2011, rach_dressing_2015}. The set of problems we consider is that of generating a specified ``target" unitary for a three-dimensional system with drift $H_0 = \mbox{diag}(0,0.515916,1)$ and control Hamiltonian with non-zero elements $\bra{1} V \ket{2} = \bra{2} V \ket{1} = 1/\sqrt{2}$ and $\bra{2} V \ket{3} = \bra{3} V \ket{2} = 1$. We restrict the control function, $\lambda(t)$, to be piecewise-constant with a specified number of segments (degrees of freedom). We first generate a set of 1000 unitary operators to use as the ``target" unitaries. We generate these by evolving the system under the Hamiltonian $H = H_0 + \lambda(t) V$ for a time $T = 0.5$ with randomly selected control functions $\lambda(t)$. We select the control functions by setting $\lambda = x_0 + x_1 t + x_2 t^2$ and choosing each of the three variables $x_0, x_1$, and $x_2$ independently and uniformly distributed on the interval $[-1,1]$. See Appendix~\ref{sec:IllustrationCoherentContr} for further discussion of these illustrations. 

\begin{figure}
    \centering
    \leavevmode\includegraphics[width = 1\columnwidth]{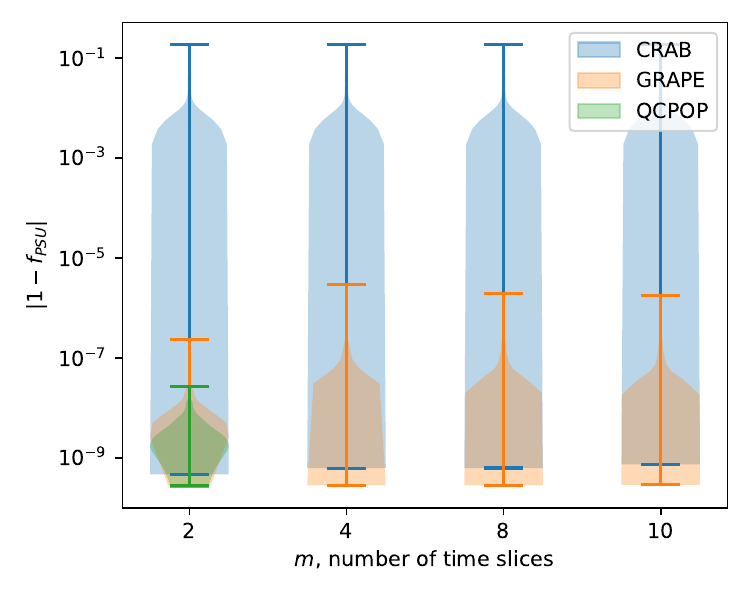}
    \caption{Violin plots of the infidelity between the target unitary, $U^\star$, and the realized unitary, $\widehat{U}$, defined in Sec.~\ref{SecEvaluatingQCPOP}, for each of 1000 target unitaries and for three quantum control methods: GRAPE, CRAB, and QCPOP. Note that $f_{PSU}$ denotes the fidelity that is defined as $f_{PSU} = \left| \Tr(\widehat{U}^\dagger U^\star ) \right| / 3$; whereas, $|1 - f_{PSU}|$ is the infidelity.} 
    \label{fig:grape_vs_pop} 
\end{figure}

We compare GRAPE, CRAB, and QCPOP on control problems defined by generating each of the 1000 unitaries in time $T = 0.5$ using 2-segment piecewise-constant control\footnote{Optimization details are provided in the Jupyter notebook at \url{https://github.com/dibondar/QControlPolyOpt/blob/master/QuantumControlPolyOpt_PiecewiseConst.ipynb}} (see also Appendix~\ref{sec:IllustrationCoherentContr}). We use QuTIP implementations of GRAPE and CRAB, and our own Julia implementation of QCPOP with the TSSOS polynomial optimization library. All three methods require approximately 1--2 seconds to find a solution. After obtaining the control field, we compute the evolution operator $\widehat{U}$ by solving Eq.~\eqref{EqPropagator}. Figure~\ref{fig:grape_vs_pop} shows the infidelity between the target unitary $U^\star$ and the synthesized $\widehat{U}$. We test CRAB and GRAPE with four initial guess types: \texttt{ZERO}, \texttt{SINE}, \texttt{LIN}, and \texttt{RND}~\footnote{See \url{https://qutip-qtrl.readthedocs.io/en/latest/apidoc/qutip_qtrl.pulseoptim.html}}. CRAB frequently becomes trapped in local minima and shows poor performance, while GRAPE performs well. However, QCPOP achieves even better results than GRAPE.

We now provide a true test of the ability of QCPOP to find global optima by applying it to a quantum control problem in which traps are so dense that it is virtually impossible to find solutions by using a gradient search method directly. This problem consists of transferring the state of one linear resonator to another by controlling the size of a linear coupling between the two. This has applications to cooling~\cite{Wang_2011} and information processing. The Hamiltonian is 
\begin{align}
    H = \hbar \omega a^\dagger a + \hbar \lambda(t) (a + a^\dagger) (b + b^\dagger) + \hbar \omega b^\dagger b , 
\end{align}
in which $\omega$ is the respective frequency of the two oscillators and $a$ and $b$ are their respective annihilation operators. Here the ``drift" Hamiltonian consists of the first and last term of $H$ and the control Hamiltonian is the second term. 

Given limits on the size (energy) of either or both of the drift and control Hamiltonians, for any non-trivial unitary there is a minimum time in which it can be realized exactly. We may wish to find a control protocol that will realize the unitary approximately in a shorter time, because the resulting reduction in decoherence during the protocol can provide an overall increase in the achieved fidelity. Placing this restriction on the duration of the protocol is one situation in which a high density of traps can appear in the search space. 

In Ref.~\cite{jacobs_fast_2016}, the authors show that for the linear state transfer problem a gradient search fails when the protocol duration $T$ is less than $\tau/3$ where $\tau = 2\pi/\omega$ is the oscillators' period. This indicates that a sharp transition occurs in the search landscape with traps becoming very dense below this point. Running 2000 searches geometrically equi-spaced between $T = \tau/100$ and $T = \tau$ achieved only a couple of solutions below $\tau/3$. Solutions can be found for all $T$ in this region using a gradient search method by augmenting it with a technique called ``path tracing", which requires repeating a gradient search many times (on the order of a thousand were used in~\cite{jacobs_fast_2016}) and, in general, it will only find local optima~\cite{MooreTibbetts,jacobs_fast_2016}. Results from the use of the above method show that there is a piecewise constant control of 5 segments that will perform an (imperfect) state transfer with an infidelity of $1\%$ in a time $T=\tau/50$. Applying QCPOP to this problem for $T=\tau/50$ with a control function with only 3 segments we find as expected that it obtains a solution in a single run. This solution also has an infidelity of $1\%$~\footnote{Details are given in the Jupyter notebook \url{https://github.com/dibondar/QControlPolyOpt/blob/master/State_transfer_in_linear\%20networks.ipynb}}. So long as the polynomial approximation used is sufficiently good, the polynomial optimization guarantees that this solution gives the best possible fidelity for $T=\tau/50$. QCPOP also shows us that there are no solutions for transfer in this time with a 2-segment control function; three-segments is the simplest possible. 

\section{Further applications} 

\subsection{Hamiltonian identification}\label{SecHI}

We have shown that the ability to write quantum control problems as polynomial optimization is very powerful. We now show that Hamiltonian identification is another problem for which polynomial optimization provides a solution. If we have a Hamiltonian $H = \sum_j c_j H_j + \lambda(t)V$ in which the parameters $\{c_j\}$ are uncertain, we can find them by: i) choosing a control function $\lambda(t)$; ii) letting the system evolve for a time $T$ and measuring the result to determine the unitary that it generates, which we will call $U^\star$. iii) The parameters $\{c_j\}$ can now be determined by minimizing $F(\{c_j\}) = \left\| U(T) - U^\star \right\|^2$ over $\{c_j\}$ where $U(T)$ is the unitary generated by $H = \sum_j c_j H_j + \lambda(t)V$. As we have shown above, this minimization problem can be well approximated by a polynomial optimization. In Appendix~\ref{sec:HamiltonianIdentification} we demonstrate this method with a numerical example. 

\subsection{Computability of quantum control problems}

As a final note, we observe that the ability to cast quantum control problems as polynomial optimization problems answers an open question in the theory of computing. In Ref.~\cite{bondar_uncomputability_2020} it was shown that the problem of determining whether a given quantum control task is possible (that is, whether there exists a set of allowed controls that will solve the problem exactly) is not computable on a Turing machine. The proof was based on the fact that digitized quantum control can solve Diophantine equations which are polynomial equations in integer-valued unknowns. The latter problem is not Turing decidable~\cite{matiyasevich_what_2011}. 

Our work here shows that if we shift the computing paradigm from that of the Turing machine, which involves discrete computations (finite precision), to that of the Blum-Shub-Smale (BSS) machine~\cite{blum_theory_1989, blum1998complexity}, which employs analog computations with an arbitrary precision, quantum optimal control problems become computable. 
Notice that the use of the BSS machine also circumvents issues \cite{gribling2023note,raghavendra2017bit,o2017sos} of bit complexity, which hinder claims of polynomial runtime in the dimensions of SDP relaxations on Turing machines.
The run-time on the BSS machine is thus determined by the rate of convergence of the Magnus expansion, which is well understood \cite{blanes2009magnus}, 
and the rate of convergence of the moment/sum-of-squares (SOS) hierarchy
(or rather its sparsity-exploiting variants), whose understanding is only very recent \cite{schlosser2024specialized,korda2025convergence}. 
We have shown here that such an approach is not only theoretically possible, but also a powerful  computational tool.

\section{Conclusion} 

Here we have shown that by formulating quantum optimal control as a polynomial optimization problem, control problems that previously required thousands of optimization runs to obtain even reasonable solutions can be solved in a single run. This tremendous increase in speed not only greatly reduces the numerical overhead of quantum control, but makes it possible to solve quantum control problems for large systems, for which it has hitherto been infeasible to perform thousands of optimizations. In addition, QCPOP is able to find the true optimal performance of quantum control by obtaining the global optimum as opposed to a sufficiently good local optimum. We have also shown that, more generally, polynomial optimization is a powerful tool and has applications in quantum technologies beyond quantum control. We expect that polynomial optimization will prove useful in other contexts in quantum engineering. 

\acknowledgments

D.I.B. is supported by the Army Research Office (ARO) (grant W911NF-23-1-0288, program manager Dr.~James Joseph) and by a cooperative agreement with DEVCOM ARL (W911NF-21-2-0139). The views and conclusions contained in this document are those of the authors and should not be interpreted as representing the official policies, either expressed or implied, of ARO or the U.S. Government. The U.S. Government is authorized to reproduce and distribute reprints for Government purposes notwithstanding any copyright notation herein. 
L.B.G. has been supported by European Union’s HORIZON–MSCA-2023-DN-JD programme under under the Horizon Europe (HORIZON) Marie Sklodowska-Curie Actions, grant
agreement 101120296 (TENORS).
J.M. and G.K. have been supported by the Czech Science Foundation
(23-07947S).
G.K. was also supported by OP
VVV project CZ.02.1.01/0.0/0.0/16\_019/0000765 ``Research Center for Informatics".

\emph{Disclaimer}.
This paper was prepared for information purposes,
and is not a product of HSBC Bank Plc. or its affiliates.
Neither HSBC Bank Plc. nor any of its affiliates make
any explicit or implied representation or warranty and
none of them accept any liability in connection with
this paper, including, but limited to, the completeness,
accuracy, reliability of information contained herein and
the potential legal, compliance, tax or accounting effects
thereof. Copyright HSBC Group 2023. 

\section*{Data Availability} 

All codes and data used in this study can be found in~\footnote{\url{https://github.com/dibondar/QControlPolyOpt}}.


\appendix

\section{Polynomial Optimization}\label{sec:PolyOptTheory}

Polynomial optimization is a fundamental problem in mathematics that frequently appears in physics. Polynomial optimization problems (POPs) involve finding the global optimum of a polynomial function subject to polynomial constraints. Mathematically, a POP can be formulated as follows:
\begin{equation}\label{P1a}
\begin{aligned}
\min_{x \in \mathbb{R}^n}  & \quad f(x) \\
\text{subject to} & \quad  g_i(x) \geq 0, \quad i=1,\ldots,m,
\end{aligned}
\end{equation}
where $f(x)$ and $g_i(x)$ are polynomials functions in $x=(x_1,\ldots,x_n) \in \mathbb{R}^n$. This problem is generally computationally hard, and finding the global optimum was considered impossible until relatively recently. The fundamental papers \cite{Lasserre2001,parrilo2003semidefinite}, building upon the work of Putinar \cite{putinar1993positive}, introduced the moment/SOS hierarchy, which provides a sequence of increasingly tighter lower bounds on the optimal value of the original polynomial optimization problem. (See \cite{monique2009} for a survey.) This results in a hierarchy $\{P_0, \ldots, P_k \}$ of semidefinite programming (SDP) problems, with $P_0$ denoting the standard moment relaxation, such that for large enough $k$ one obtains the global optimum (although in certain problems this requires $k \to \infty$). 

Let us formalize the above by considering the POP \eqref{P1a} that we refer to as $P$. In order to construct the moment/SOS hierarchies, we need to define the basic semi-algebraic set
$K = \{x \in   \mathbb{R}^n \,\,|\,\,g_j(x) \geq 0 \}$,
such that $P$ can be reformulated as
\begin{align}\label{P1b}
    P: \quad f^* &= \min \, \{ f(x) \,\,| \,\, x \in K \}.
\end{align}
Then let us define the family of problems
\begin{align}\label{Pglobal0}
    P_\lambda: \quad f^* = \sup \{\lambda \,\,|\,\, f(x)-\lambda \geq 0, \quad \forall x\in K \ \}.
\end{align}
 Lass\'ere \cite{Lasserre2001} and Parrilo \cite{parrilo2003semidefinite} provided a systematic algorithmic procedure to construct certificates of positivity on $K$ for $f(x)-\lambda$ which are then used to obtain the solution of Prob. \eqref{Pglobal0} in practice. 

Concretely, Putinar's Positivstellensatz states the following: If a real polynomial $f$ is strictly positive on $K$ then it can be written as 
\begin{align}\label{SOS}
    f(x) = \sigma_0(x) + \sigma_1(x)g_1(x) + \ldots \sigma_m(x)g_m(x),
\end{align}
for any real vector $x$, where $\sigma_j$ are real SOS polynomials. Testing whether $f(x)$ can be written as in Eq. \eqref{SOS}, signifying the desired result $f(x)>0$ for any $x\in K$, amounts to solving an SDP. The dual of this SDP, which features prominently in the approach of Lassere \cite{Lasserre2001}, is the problem whether given a real sequence $y = \{ y_\alpha\}$ with $\alpha \in \mathbb{N}^n$ a multi-index, if there exists a probability measure $\mu$ on $K$ such that
\begin{equation}
    y_\alpha = \int_K x_1^{\alpha_1}\ldots x_n^{\alpha_n}\, d\mu,
\end{equation}
for any $\alpha$, then $y$ represents a measure supported on $K$. Then, one introduces a linear function $L_y$ that maps real polynomial functions to real numbers defined as
\begin{align}
    f = \sum_{\alpha \in \mathbb{N}^n}f_\alpha x^\alpha \mapsto L_y(f) = \sum_{\mathbb{N}^n}f_\alpha y^\alpha.
\end{align}
The sequence $y$ above has a representing measure on $K$ if and only if for every real polynomial $h$ it holds that  
\begin{align}\label{condition}
    L_y(h^2) \geq 0, \quad L_y(h^2g_j) \geq 0,
\end{align}
for $j=1,\ldots,m$. If Eq. \eqref{condition} holds for all polynomials $h$ with maximum degree $d$, certain $m+1$ moment and localizing matrices, with entries linear in $\{ y_\alpha\}$, are positive semidefinite:
\begin{align}\label{moments}
    M_d(y) \succeq 0, \quad M_d(g_jy) \succeq 0,
\end{align}
for  $j=1,\ldots,m$. The latter problem, Prob. \eqref{moments}, can be modeled as a so-called ``generalized moment problem'':
\begin{align}
    \inf_{\mu_i} \left\{ \sum_{i=1}^\ell \int_{K_i} f_i \, d\mu_i \,\,\Bigg| \,\, \sum_{i=1}^\ell h_{ij}\, d\mu_i \geq b_j  \right\},
\end{align}
where $K_i \subset \mathbb{R}^{n_i}$ is defined as before.

From either point of view, the moment/ SOS approach amount to solving a hierarchy of (convex) semidefinite programs of increasing size:
\begin{align}
    f_d^* = \sup_{\lambda,\sigma_j} \left\{ \lambda \,\, \Bigg| \,\, \sigma_0 + \sum_{j=1}^m \sigma_jg_j, \text{ with } {\rm deg}(\sigma_jg_j) \leq 2d \right\} 
\end{align}
The sequence $\{f_d^* \}$, where $d\in \mathbb{N}$, is monotone and non-decreasing. When $K$ is compact, one obtains the global optimum of Prob. \eqref{Pglobal0} as the limit
\begin{align}
    f^* = \lim_{d\to \infty} f_d^*,
\end{align}
where finite convergence is generic. 


\subsection*{Exploiting sparsity}

Polynomial optimization over a basic semialgebraic set is NP-hard \cite{monique2009}. While the moment/SOS  hierarchy of relaxations, which was discussed above, establishes a framework for tackling such
problems, they remain hard and despite the popularity and power of the framework, certain limitations to the scalability of the moment/SOS  hierarchy  exist. Specifically, for the number $n$ of variables of the POP at hand, the size of the matrices appearing in the $d$-th step of the moment/SOS relaxation is of the order of  $\binom{n+d}{n}$. 

To address this challenge, several proposals have been put forth. Putinar \cite{putinar1993positive} suggested removing monomials not appearing in the moment/SOS decomposition. 
Lasserre \cite{Lasserre2006} suggested exploiting possible sparsity patterns observed in the polynomial optimization problem. 
In some sense, this amounts to re-indexing the SDP matrices
involved in the moment/SOS relaxation as follows: consider subsets $I_1, \ldots , I_p \subseteq \{1, . . . , n\}$ of the initial input SDP variables, with cardinalities $\{|I_1|, \ldots, |I_p|\}$ respectively. The sparse moment/SOS hierarchies \cite{Lasserre2006,TSSOS,ChordalTSSOS}, consider (quasi) block-diagonal SDP matrices where the size of block $j$ is given by $|I_j|$, $1\leq j \leq p$. With this reformulation of the original SDP, if the cardinalities above are small compared to the initial variables, that is if $|I_j| \ll n$ for all $j$, applying the moment/SOS relaxations achieves significant computational savings, which in turn  enables scalability. Under mild assumption, theoretical guarantees on the global convergence of the sparse moment/SOS relaxations exist. 

\subsection{Numerical algebraic geometry}\label{sec:HomotopyContinuation}

An alternative approach leverages a large body of work within numerical algebraic geometry \cite{lindberg2023polyhedral}.
Let us consider the polynomial optimization problem \eqref{P1a} and its Lagrangian:
\begin{align}
    L(x, \lambda) := f(x) + \sum_{j=1}^{m} \lambda_j g_j(x).
\end{align}
The first-order optimality conditions, known as the Lagrange system 
\begin{align}
    F(x, \lambda) := \left\{\pfpx{L(x, \lambda)}{x_i}~\Bigg| ~i=1,\dots, n\right\},
\end{align}
is a system of polynomial equations, whose solutions (``roots'') are sought in numerical algebraic geometry. 
If one could enumerate all roots of the algebraic system, one could find the global optimum
by a linear scan. 

To introduce the so-called homotopy-continuation approach \cite{sommese2005numerical,allgower1997numerical,burgisser2013condition} within numerical algebraic geometry, one can consider the so-called straight-line homotopy.
There, we see $F$ as the ``target system'', pick a suitable ``start system'' $G$ that has the same or higher 
number of roots, known to us, and consider a homotopy:
\begin{align}
    H(x,\lambda, t) = tF(x,\lambda) + (1-t)G(x,\lambda).
\end{align}
As one traces the path $t \in [0, 1]$ for all roots of the start system, one may be able to arrive
at all the roots of the target system. 
Traditionally, the method is presented in terms of complex-valued polynomials, but one can clearly
filter for real-valued roots. 
While this may seem straightforward, several challenges have made numerical algebraic geometry into an active area of research over the past three decades. 

The first challenge is related to the scalability and exploitation of a sparsity in the system. 
To some extent, this can be addressed by the choice of the start system. The theorem of 
Bernstein \cite{Bernstein75,Khovanski78,Kushnirenko76}, also known as the Bernstein–Khovanskii–Kushnirenko (BKK) theorem, provides 
an upper bound on the number of isolated solutions a sparse polynomial system, which is exact, 
generically.   
This underlies the polyhedral homotopy algorithm of Huber and Sturmfels \cite{huber1995polyhedral}. 
\texttt{HomotopyContinuation} utilizes the BKK theorem to bound the number of roots of the target
system and pick the matching start system.

The second challenge is related to numerical stability of tracing the paths (``path-tracking'') from the start
system to the target system.
As a path approaches a singular endpoint, one needs to modify the continuation process
to improve numerical stability. 
Morgan et al.\ \cite{morgan1990computing,morgan1992computing,morgan1992power} coined the term ``end-game'' 
for approaches based on uniformization theorem for germs of one-dimensional analytic sets, which have been 
subsequently extended by numerous other researchers \cite[e.g.]{huber1998polyhedral}.
We employ the end-game based on power series \cite{morgan1992power} and Cauchy \cite{morgan1992computing} endgame, as implemented in \texttt{HomotopyContinuation}~\cite{HomotopyContinuation.jl}.

The third challenge is related to the post-hoc certification of the correctness and completeness of the set of roots. 
Leading approaches in this direction \cite{hauenstein2012algorithm,breiding2020certifying} are
based on the point-estimation theory of Smale (\cite{smale1986newton}, also known as the $\alpha$-theory) and interval arithmetics \cite{moore1977test}, respectively. 
\texttt{HomotopyContinuation} implements a technique of \cite{breiding2020certifying}
based on interval arithmetics. 

\section{Illustrations}\label{sec:IllustrationCoherentContr}

\begin{figure}
    \centering
    \includegraphics[width=0.45\textwidth]{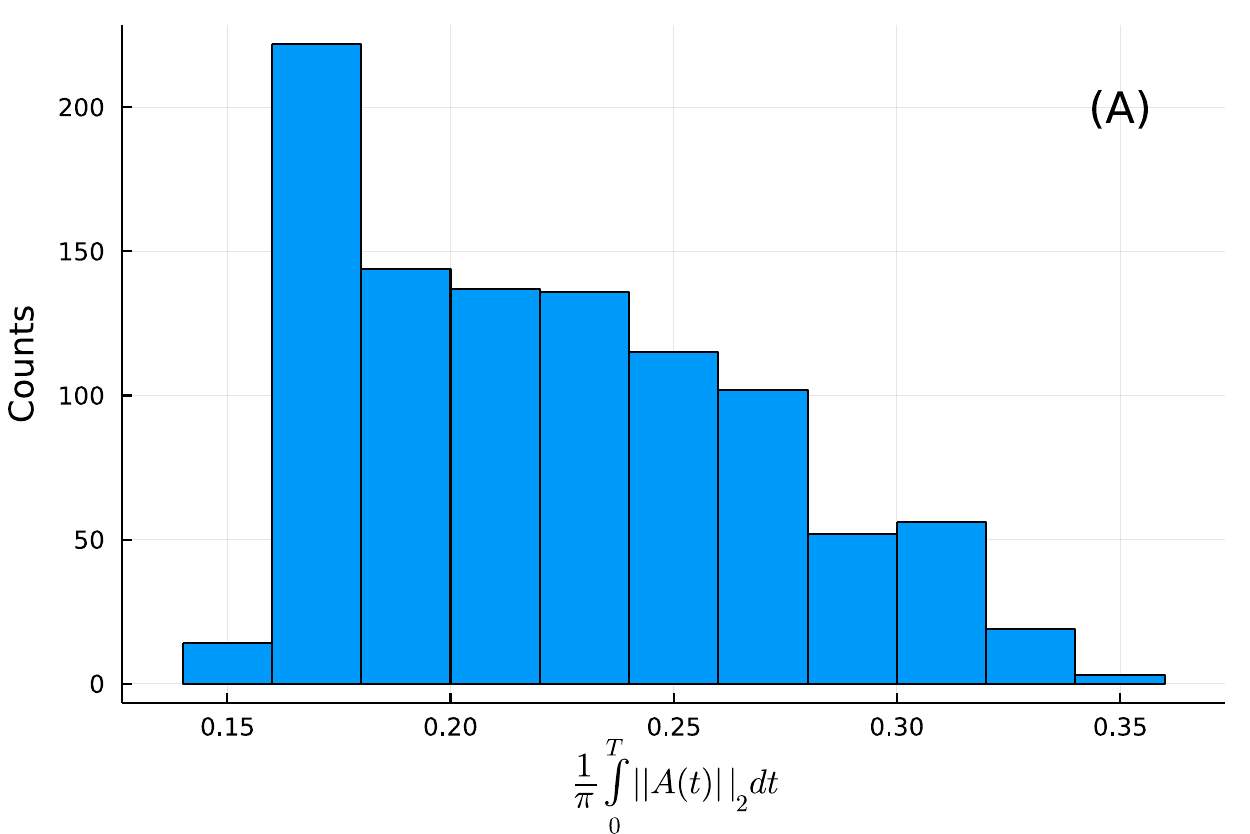}
    \includegraphics[width=0.45\textwidth]{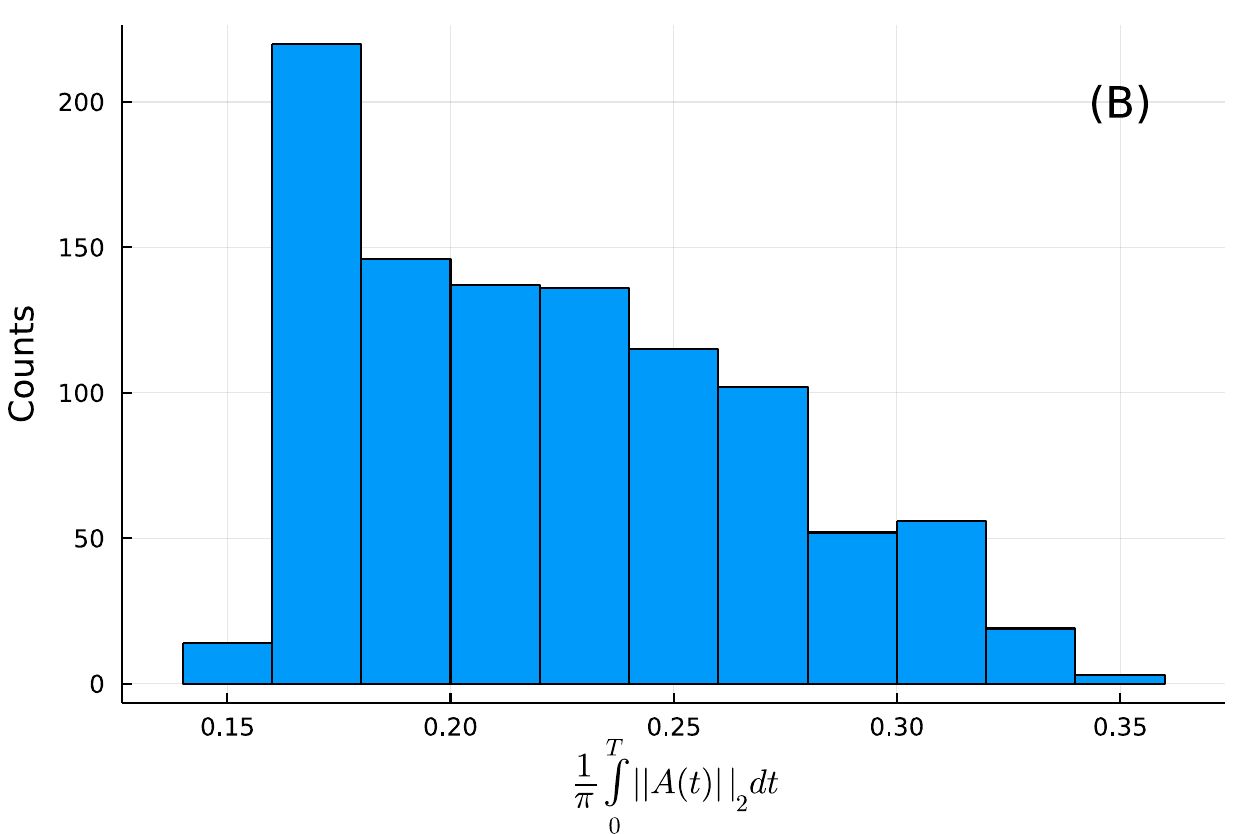}
    \caption{Histogram plots for Magnus series convergence tests [see the main text just after  Eq.~\eqref{eq:Omega3}] (A) when exact solutions $\bm x^\star$ for quantum control problem~\eqref{eq:QuantumCoherentControl} are used; (B) when global minimizers $\widehat{\bm x}$ for~\eqref{Eq:UminusUtarget}, obtained by the \texttt{TSSOS} Julia package, are employed.}
    \label{fig:ConvergeneTest}
\end{figure}

\begin{figure}
    \centering
    \includegraphics[width=0.5\textwidth]{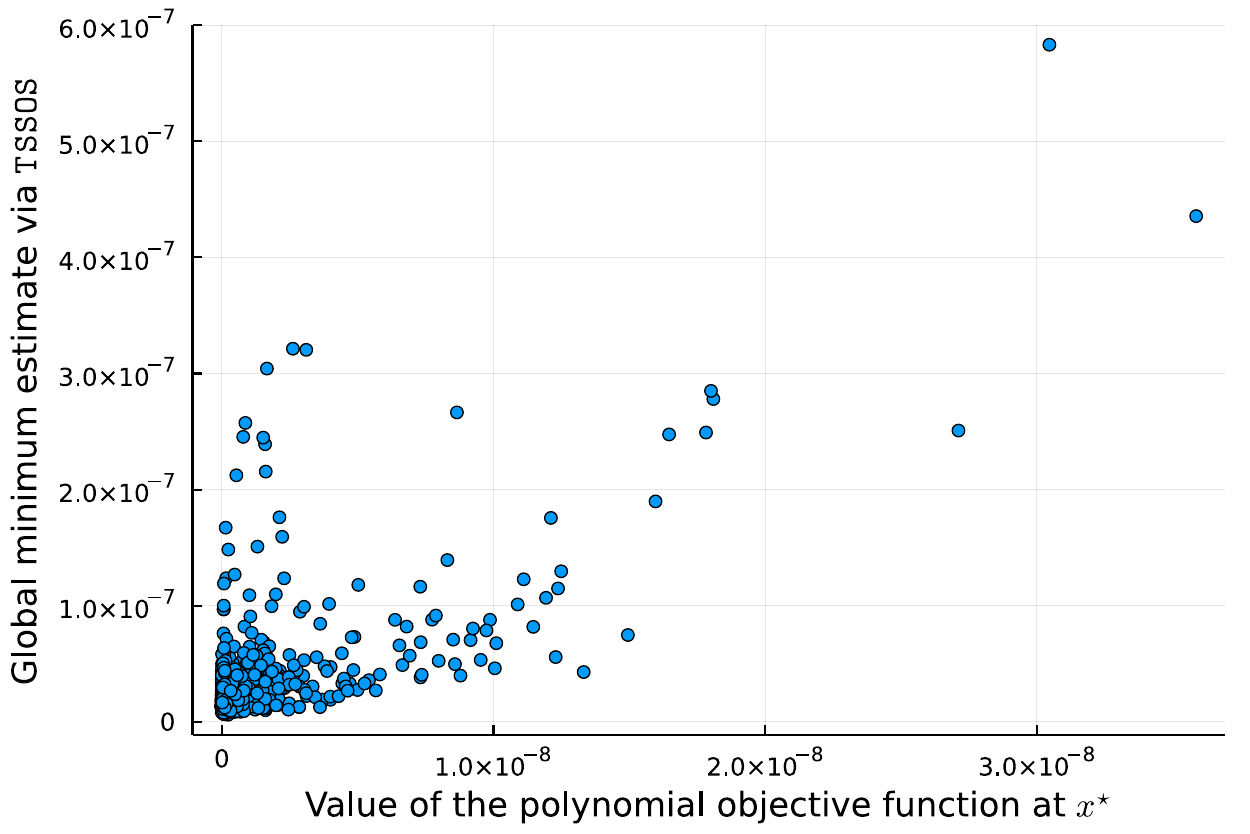}
    \caption{Comparing the global minima for~\eqref{Eq:UminusUtarget}, obtained by the \texttt{TSSOS} Julia package, with the value of the polynomial objective function~\eqref{Eq:UminusUtarget} when an exact solution $\bm x^\star$ for quantum control problem~\eqref{eq:QuantumCoherentControl} is used.}
    \label{fig:TSSOSvsXstar}
\end{figure}

\begin{figure}
    \centering
    \includegraphics[width=0.5\textwidth]{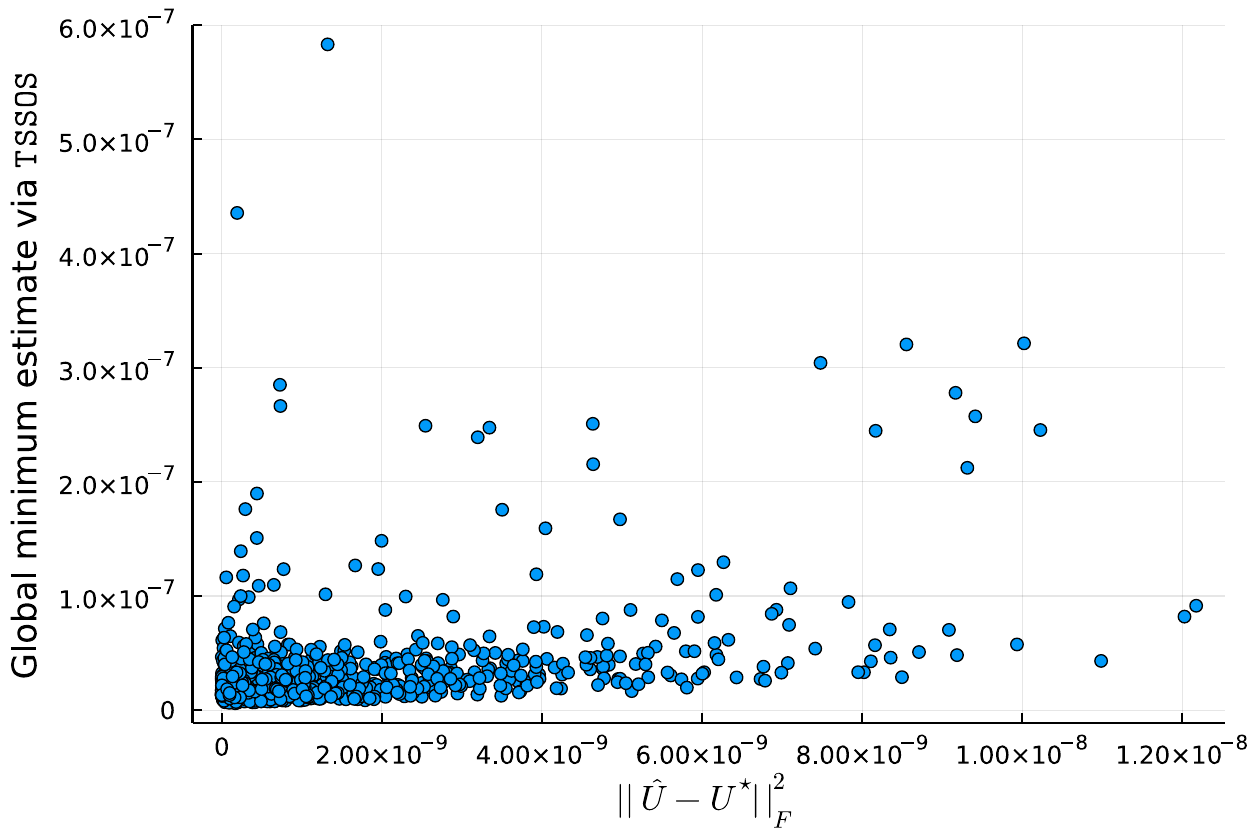}
    \caption{A relation between the quantum control problem~\eqref{eq:QuantumCoherentControl} and its approximate polynomial formulation~\eqref{Eq:UminusUtarget}. $\widehat{U}$ is the evolution operator obtained via polynomial optimization~\eqref{Eq:UminusUtarget}.}
    \label{fig:PolynomialOptVsCoherenControl}
\end{figure}

In this section, we provide numerical illustrations for the polynomial  formulation~\eqref{eq:QuantumCoherentControl} of the terminal quantum control problem of synthesizing a desired unitary gate. The code used in the analysis is available as a Julia Jupyter notebook~\footnote{\url{https://github.com/dibondar/QControlPolyOpt/blob/master/QuantumControlPolyOpt.ipynb}}. 
In particular, we utilize the drift and control Hamiltonians of the form
\begin{align}\label{eq:QuantumExample}
    H_0 = \begin{pmatrix}
        0 & 0 &  0 \\
        0 & 0.515916 & 0 \\
        0 &  0 &       1 
    \end{pmatrix},
    \hfill 
    V = \begin{pmatrix}
        0 & 0.707107 & 0 \\
        0.707107 & 0 & 1 \\
        0 & 1 & 0
    \end{pmatrix},
\end{align}
which are scaled versions of the Hamiltonians for IBM Q 2-qubit devices~\footnote{\url{https://github.com/q-optimize/c3/blob/master/examples/two_qubits.ipynb}}. The analysis is started by randomly drawing components of the vector $\bm x^\star$ ($m = 3$) from a uniform distribution on the interval $[-1, +1]$. In total, 1000 samples are generated. Then, for each  sample, we perform the following steps: Setting $\bm x = \bm x^\star$ in Eq.~\eqref{eq:E2x}, we obtain the control field used to get the target unitary $U^\star$ by solving the Schr\"odinger equation~\eqref{Eq:Schrodinger}. The terminal time is chosen at $T=0.5$. Hence, we know that $\bm x^\star$ is an exact solution to the terminal quantum control problem~\eqref{eq:QuantumCoherentControl}. Figure~\ref{fig:ConvergeneTest}(A) shows that the inequality $ 
    \int _{0}^{T}\| A(t) \|\, dt < \pi$ is satisfied [see the main text just after Eq.(\ref{eq:Omega3})], therefore justifying the usage of the truncated Magnus series to approximate the evolution operator. 
The polynomial formulation~\eqref{Eq:UminusUtarget} with $n=3$ and $p=5$ is then solved using the \texttt{TSSOS} Julia package~\cite{TSSOS,ChordalTSSOS}. This library extracts the minimizer $\widehat{\bm x}$, i.e., the value of $\bm x$ yielding the global minimum estimate for~\eqref{Eq:UminusUtarget}. It is noteworthy that the obtained global minimum estimate via \texttt{TSSOS} is larger than the value of the objective function~\eqref{Eq:UminusUtarget} at $\bm x^\star$ (see Fig.~\ref{fig:TSSOSvsXstar}) as expected from the monotone convergence property of the hierarchy of SDP relaxations (see Sec.~\ref{sec:PolyOptTheory}). Hence, the polynomial optimization yields $\widehat{\bm x} \neq \bm x^\star$. Notice also the well-known fact that quantum control problems typically have non-unique solutions~\cite{rabitz_quantum_2004, brif_control_2010}. 
Figure~\ref{fig:ConvergeneTest}(B) shows that the Magnus expansion for $\widehat{\bm x}$ converges. The control field [Eq.~\eqref{eq:E2x}] obtained from $\widehat{\bm x}$ generates the evolution operator $\widehat{U}$ that is supposed to be very close  to the target $U^\star$. Figure~\ref{fig:PolynomialOptVsCoherenControl} indeed confirms that. 

We conclude that the approximate polynomial formulation~\eqref{Eq:UminusUtarget} provides a viable alternative for the original quantum control problem~\eqref{eq:QuantumCoherentControl}.

\section{Hamiltonian Identification}\label{sec:HamiltonianIdentification}

Let us now consider an easy starting point in this direction. For convenience, assume that we know the drift Hamiltonian $H_0$ \eqref{eq:QuantumExample}, but we do not know the values of nonzero coupling constants in the interaction Hamiltonian $V$ \eqref{eq:QuantumExample}. Hence, $V$ is assumed to be of the form
\begin{align}\label{eq:V_unknown}
     V = \begin{pmatrix}
        0 & z_1 & 0 \\
        z_1 & 0 & z_2 \\
        0 & z_2 & 0
    \end{pmatrix},
\end{align}
and our goal is to find the unknown coupling constants $\bm z = (z_1, z_2)$.
We note that this a physically important formulation of the problem of Hamiltonian parameter identification. The diagonal elements correspond to eigenenergies of the non-driven quantum system, which are easy to measure, and thence can be assumed to be known accurately. The positions of zeros elements in the interaction Hamiltonian  $V$ \eqref{eq:QuantumExample} follows from the symmetry of physical systems, 
e.g., via the Wigner--Eckart theorem.

To determine the unknown coupling constants $\bm z$, we drive the quantum system 
with a known control signal $\bm x$, repeatedly, and perform the tomography at the end of the evolution to determine the synthesized unitary $U^{\star}$. 
Experimentally, we show that it is often sufficient to know only a single pair $\bm x \to U^{\star}$, in order to very accurately recover the unknown couplings $\bm z$ by solving the following unconstrained polynomial optimization problem:
    \begin{mini}|l|
    {\bm{z}}{ \left\| e^{\Lambda_n/2} -  e^{-\Lambda_n/2} U^\star \right\|^2}{}{}.
    \label{eq:POPHI}
    \end{mini}
The formulation~\eqref{eq:POPHI} is a minor modification of the polynomial optimization of quantum control \eqref{Eq:UminusUtarget}. The difference lies in the fact that in the latter, the control signal $\bm x$ is to be found, using a known Hamiltonian (i.e., known $\bm z$). In the former \eqref{eq:POPHI}, we wish to find parameters of the interaction Hamiltonians, using the known control signal.
\begin{figure}
    \centering\includegraphics[width=0.5\textwidth]{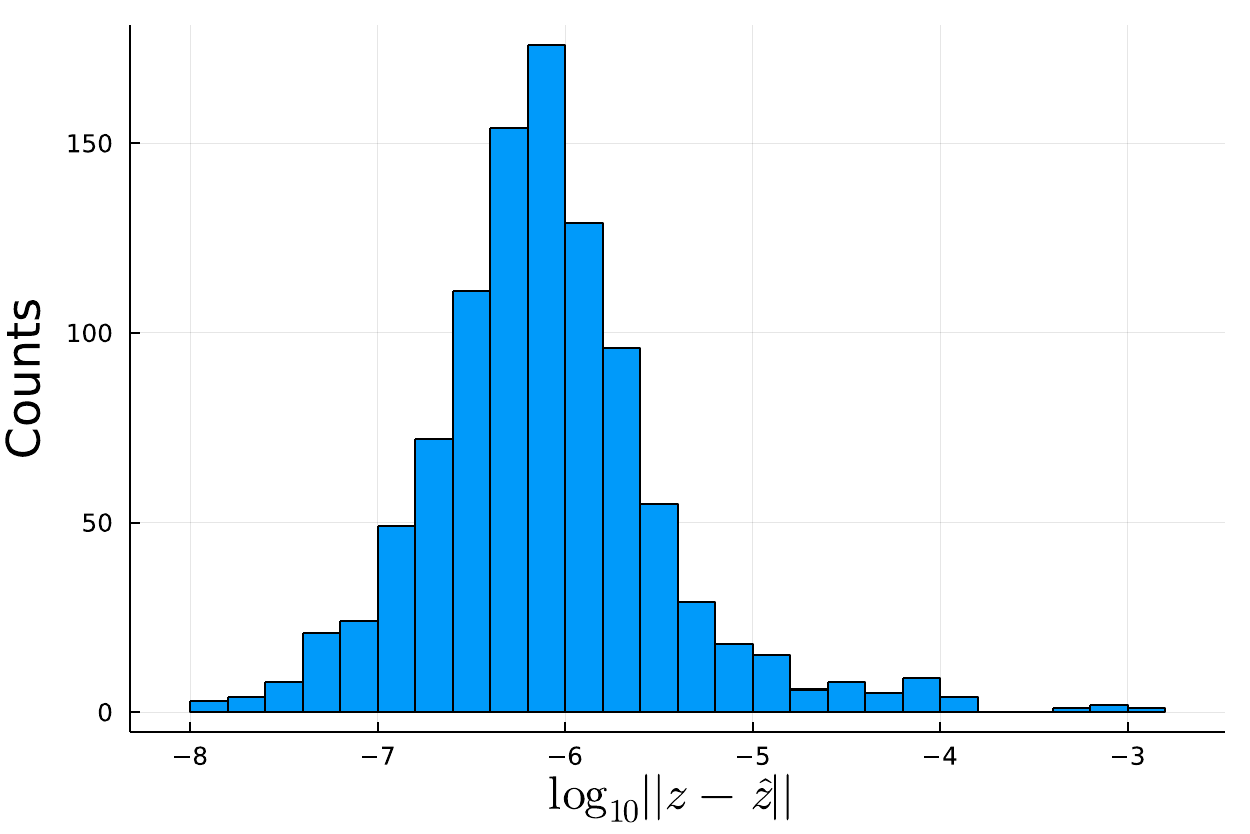}
    \caption{An illustration of the effectiveness of polynomial optimization formulation~\eqref{eq:POPHI} for the problem of Hamiltonian identification. Distribution of the error between the exact $\bm z$ and recovered $\widehat{\bm z}$ non-zero element of the interaction Hamiltonian~\eqref{eq:V_unknown}.}
    \label{fig:ZexactVsZrecovered}
\end{figure}

To numerically illustrate the effectiveness of formulation~\eqref{eq:POPHI}, we perform an analysis similar to that of Sec.~\ref{sec:IllustrationCoherentContr}. We start by randomly drawing components of the vector $\bm x$ ($m = 3$) from a uniform distribution on the interval $[-1, +1]$. In total, 
1000 samples are generated. Then, for each sample, we perform the following steps: We obtain the terminal unitary $U^\star$ at $T=0.5$ by solving the Schr\"odinger equation~\eqref{Eq:Schrodinger} 
using the exact Hamiltonian \eqref{eq:QuantumExample}. The polynomial minimization~\eqref{eq:POPHI} with $n=3$ and $p=5$ is solved via the \texttt{TSSOS} Julia package \cite{TSSOS,ChordalTSSOS} yielding the estimate $\widehat{\bm z}$ of the global minimizer for $\bm z$. The obtained estimate is then refined by performing a local minimization. The code used for this analysis can be found in the accompanying Julia Jupyter notebook~\footnote{\url{https://github.com/dibondar/QControlPolyOpt/blob/master/HamiltonianParameterIdentification.ipynb}}. The results are shown in Fig.~\ref{fig:ZexactVsZrecovered}, where the distribution of the errors between the exact couplings $\bm z$ and recovered couplings $\widehat{\bm z}$ are shown for all samples. As can be readily seen, the Hamiltonian is recovered with a very high accuracy even from a single pair $\bm x \to U^{\star}$. 

\bibliography{literature}

\end{document}